\documentclass[preprintnumbers,showpacs,aps,amsmath,amssymb,superscriptaddress,nofootinbib,prl,twocolumn,10pt]{revtex4-1}
\usepackage{epsfig}
 \usepackage{euscript}
\usepackage{amssymb}
\usepackage{amsfonts}
\usepackage{amsbsy}
\usepackage{amsmath}
\usepackage{epsfig}
\usepackage{amsthm}
\usepackage{amscd}
\usepackage{amstext}
\usepackage{verbatim}
\usepackage{amsfonts}
\usepackage{graphicx}
\usepackage{epsfig}
\usepackage{eepic}
\usepackage{amsmath}
\usepackage{amssymb}
\usepackage{enumerate}
\usepackage{color}
\usepackage{dcolumn}
\usepackage{bm}
\usepackage{ulem}

\def\bc{\begin{center}}
\def\nno{\nonumber}
\def\ec{\end{center}}
\def\be{\begin{eqnarray}}
\def\ee{\end{eqnarray}}
\definecolor{dyellow}{rgb}{1.,0.8,.0}
\definecolor{myblue}{rgb}{.1,.1,.7}
\definecolor{dcyan}{rgb}{.0,.6,.6}
\definecolor{dmagenta}{rgb}{0.6,0.0,0.6}
\definecolor{brown}{rgb}{0.6,0.2,0.}
\definecolor{darkblue}{rgb}{.0,.0,0.5}
\definecolor{darkred}{rgb}{0.75,0.0,0.0}
\definecolor{orange}{rgb}{1.,.6,.0}
\definecolor{dorange}{rgb}{0.8,.4,.0}
\definecolor{darkgreen}{rgb}{0.0,0.6,0.0}
\definecolor{purple}{rgb}{.4,.0,.4}
\definecolor{lightgrey}{rgb}{0.7, 0.7, 0.7}
\definecolor{grey}{rgb}{0.4, 0.4, 0.4}

\newcommand{\nc}{\newcommand}
\nc{\rnc}{\renewcommand} \nc{\ket}[1]{\left | \, #1 \right \rangle}
\nc{\bra}[1]{\left \langle #1 \, \right |}
\nc{\ua}{\uparrow} \nc{\da}{\downarrow}

\nc{\braket}[2]{\langle\, #1\,|\,#2\,\rangle}
\nc{\half}{\frac{1}{2}}

\nc{\prj}{\mathcal{P}} \nc{\hilb}{\mathcal{H}}
\nc{\pth}{\mathcal{C}} \nc{\inprod}[2]{\braket{#1}{#2}}
\nc{\upket}{\ket{\uparrow}} \nc{\downket}{\ket{\downarrow}}
\nc{\upbra}{\bra{\uparrow}} \nc{\downbra}{\bra{\downarrow}}

\begin{document}
\title{Possible Anderson localization in a holographic superconductor}

\author{Hua Bi Zeng}
\affiliation{School of Mathematics and Physics, Bohai University
JinZhou 121000, China}
\affiliation{CFIF, Instituto Superior T\'ecnico,
Universidade T\'ecnica de Lisboa, Av. Rovisco Pais, 1049-001 Lisboa, Portugal}

\begin{abstract}
We study the effect of disorder in a holographic superconductor by introducing a quasi-periodic chemical potential.
When the condensation of the superconductor is sufficiently small compared with the strength of disorder, we find that there exists a discontinuous phase transition from superconducting state to normal state with increasing disorder strength.
For relatively large condensation,
we find that disorder suppress but not completely destroy superconductivity.

\end{abstract}
\maketitle
\pagebreak

{\it Introduction.-}
The effect of disorder in superconductor has intrigued scientists for several
decades. Soon after the BCS theory \cite{1}, Anderson found that weak disorder  cannot destroy the superconductivity \cite{2}. Until now, both theories and experiments have confirmed that a strong disorder will eventually destruct superconductivity, driving the system into an insulating state or a normal metal state \cite{3,31,32,33,35,36,37,38}.
However, the effect of interactions in a disordered superconductors is still not well understood.
As a natural way to study a strongly coupled quantum field theory systems,
the AdS/CFT correspondence \cite{adscft} has been used to study the interplay of disorder and interaction \cite{4,9,5,6,7,8}. The holographic correspondence has also been proved to
be successful to study various properties of superconductors\cite{h1,h2}.
In reference \cite{d1} the authors firstly studied a dirty holographic superconductor,
the found that the disordered superconductor always has a larger critical temperature relative to the to
the $T_c$ for the uniform one.
In this paper we focus on understanding another important issue, the
possible Anderson localization in a holographic superconductor.
\textcolor{blue}{Technically, the weak disorder effect is introduced by a quasi-periodic
chemical potential on the boundary field theory, the strength of the disorder is controlled
by a parameter $\alpha$. By tuning $\alpha$ we find that when the
condensation is small, the weak disorder will destroy the superconductivity,
clearly this is a holographic realization of Anderson localization in superconductors. }

{\it Model and definition of disorder.-}
The starting action in the usual gravity dual of a holographic superconductor is \cite{h1}
$
S=\int d^4x\sqrt{-g}[R-2\Lambda-\frac14F_{\mu\nu}F^{\mu\nu}-|\nabla\psi-iA\psi|^2-m^2|\psi|^2]$'
where $\Lambda=-d(d-1)/2L^2$ is the cosmological constant, $d$ is the dimension of the boundary, and $F_{\mu\nu}=\partial_\mu A_\nu-\partial_\nu A_\mu$ is the strength of the gauge field. The metric is an AdS Schwarzschild black hole,
$ds^2=-f(r)dt^2+\frac{dr^2}{f(r)}+r^2(dx^2+dy^2)$
with $f(r)=r^2/L^2(1-r_0^3/r^3)$, $r$ being the bulk radial coordinate, $r_0$ the horizon position, and $x,y$ the boundary coordinates. Without loss of generality, we set $ L=1$.
The temperature of the black hole is $T=\frac{3 r_0}{4 \pi}$.

We use the ansatz of $\psi=\psi(r,x)$ and $A=(A_t(r,x),0,0,0)$, where
$x$ is the spatial coordinate of the boundary field theory,
and choose $m^2=-2$.
In the probe limit, with the scaling of $\psi\rightarrow \psi/r$ and working in the
coordinates with $z=1/r$, we have the following equations of motion (EoMs):
\begin{widetext}
\be
&&\left(1-z^3\right) A_t^{(2,0)}(z,x)+ A_t^{(0,2)}(z,x)-2
    A_t(z,x) \psi (z,x)^2=0,\\
   &&\psi (z,x) \left( A_t(z,x)^2+z^4-z\right)+\left(1-z^3\right)
   \psi ^{(0,2)}(z,x)+\left(z^3-1\right)^2 \psi ^{(2,0)}(z,x)\nno\\&&+3
   \left(z^3-1\right) z^2 \psi ^{(1,0)}(z,x)=0.
\ee
\end{widetext}

\begin{figure}
\begin{center}
\includegraphics[trim=0cm 9.5cm 0cm 9.5cm, clip=true,scale=0.45]{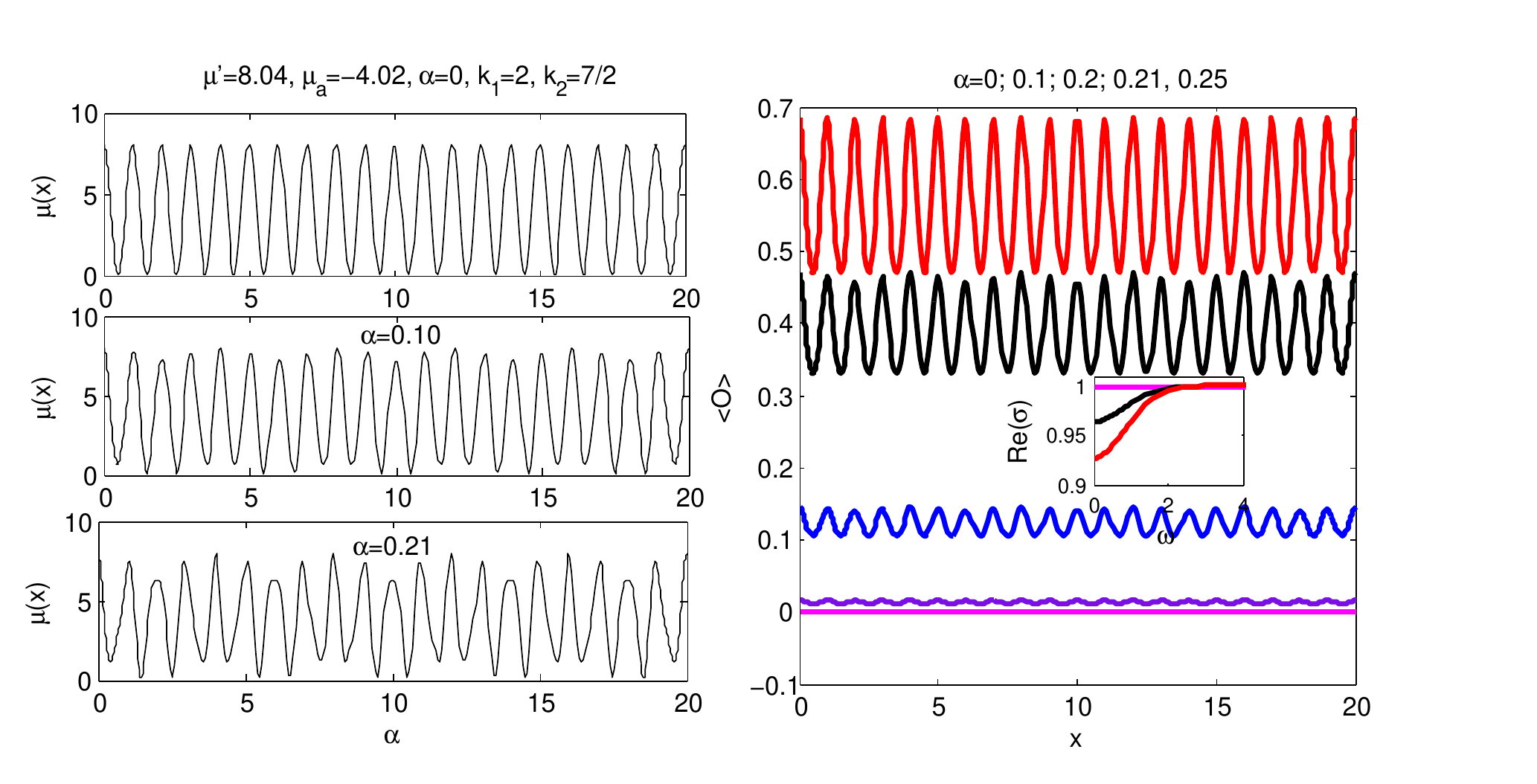}
\caption{left: The plot of $\mu(x)$ for the cases of $\mu_a=4.02, \mu'=8.04$ and $\alpha=0; 0.1; 0.21$ with $k_1=2, k_2=7/2$. right:
The order parameter $\langle \mathcal O(x) \rangle$ for five cases $\alpha=0; 0.1; 0.2; 0.21; 0.25$ (from top to bottom) with $\mu_a=4.02, \mu'=8.04, k_1=2, k_2=7/2$, we see that a phase transition happens when increasing $\alpha$. Inset of  the right plot:
The average values of real part of conductivity along the $y$ direction for three cases $\alpha=0; 0.1; 0.25$ in the right plot.}\label{fig1}
\end{center}
\end{figure}

The superscripts on the
fields mean the derivative of $z$ and $x$, for example $A_t^{(2,0)}(z,x)$
means $\partial_z^2  A_t(z,x)$ and $A_t^{(0,2)}(z,x)$
means $\partial_x^2  A_t(z,x)$.
The expansions of $\psi$ and $A_t$ near the infinite boundary are:
\be\label{eomr}
\psi(r,x)&\sim&\psi^{(0)}(x)+\psi^{(1)}(x) z+\dots,\\
A_t(r,x)&\sim&\mu(x)+\rho(x) z+\dots.
\ee
We choose the quantization such that $\psi^{(0)}(x)=0$ and $\psi^{(1)}(x)=\langle \mathcal O(x) \rangle$ is the order parameter.
We introduce the disorder through a  quasi-periodic chemical potential on the boundary as:
\be
\begin{split}
\mu(x)=\mu_a+(1-\alpha)(\mu'-\mu_a)\cos(2 k_1\pi x /2)+\\
\alpha (\mu'-\mu_a)\cos(2 k_2\pi x/2 ),
\end{split}
\ee
where $2 k_1$ and $2 k_2$ are two coprime positive integers,
$\mu_a$ is the average value of $\mu(x)$,
$0\leq\alpha\leq 1$ controls the pattern of $\mu(x)$,
$\mu'$ controls the maximal value $\mu_{max}=\mu'$ of $\mu(x)$ and the minimal value $\mu_{min}= -\mu'+2 \mu_a$ of $\mu(x)$.
Thus, the amplitude of the oscillating $\mu(x)$ is $2(\mu'-\mu_a)$.
Then $\alpha$ is the parameter of  the disorder strength after fixing $\mu_a$ and $\mu'$,
$\mu'$ is the parameter of the amplitude of the oscillation after fixing $\mu_a$ and $\alpha$.
Similar kind of quasi-periodic lattice has been already used to study the Anderson localization in Refs. \cite{qp1,qp2,qp3}.
 Strictly, $k_1 / k_2$ should an irrational number for the quasi-periodic case,
however, by using two coprime positive integers $2 k_1$ and $2 k_2$ we can still
induce some weak disorder effect as shown in Fig. \ref{fig1}.

The EoMs are solved by using the Chebyshev spectral method \cite{spectral}.
We discretize the EOMs on a two dimensional Chebyshev grid with $20$ points along the
$z$ direction and $400$ points in the $x$ direction. In all the calculations we choose the length $l$ of the sample to be $l=20$.




\begin{figure}
\begin{center}
\includegraphics[trim=0cm 9.2cm 3cm 11.2cm, clip=true,scale=0.55]{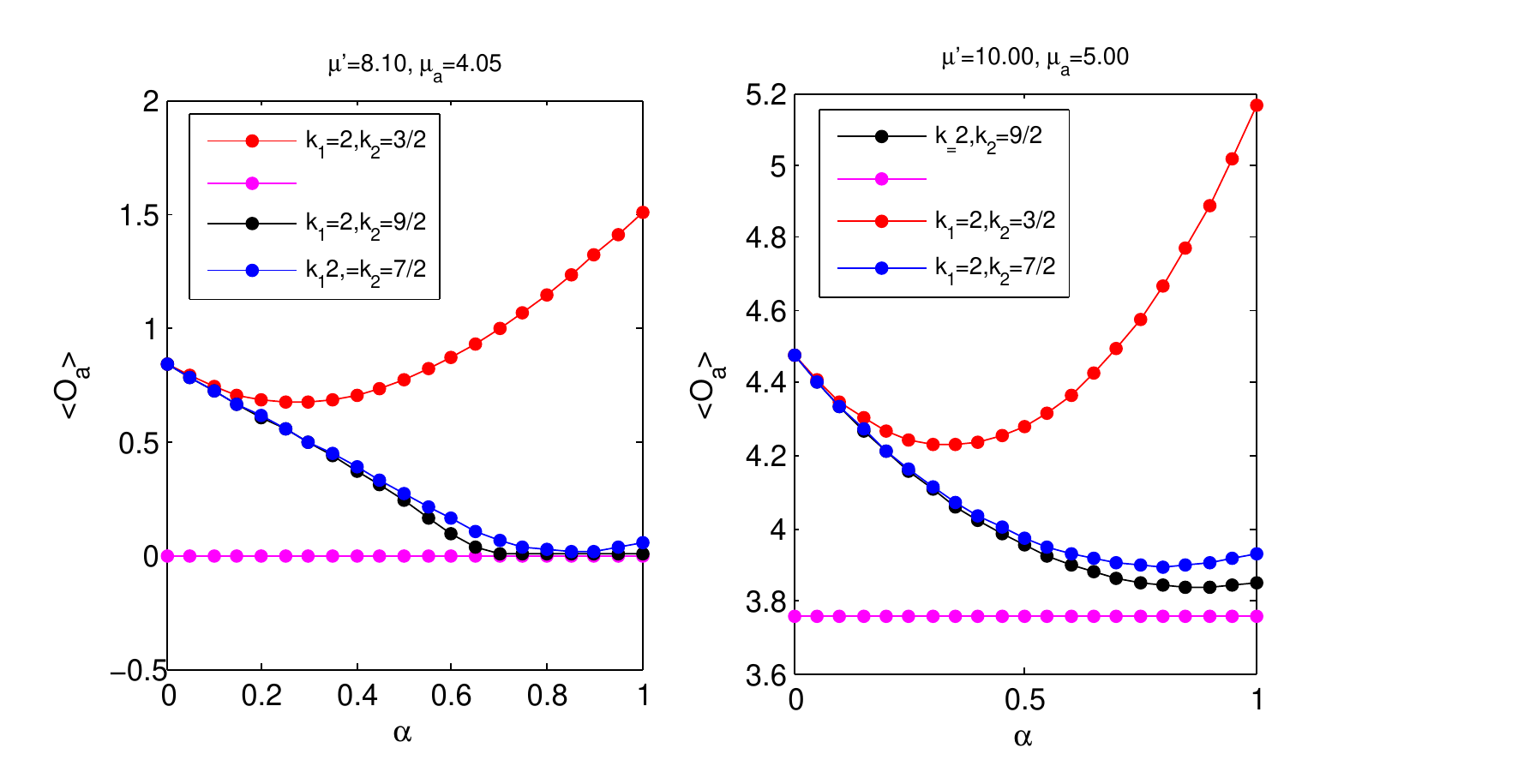}
\caption{The average value of order parameter $\langle \mathcal O_a \rangle$ as a function of $\alpha$ for different $k_2$ with a fixed $k_1=2$. In The left plot $\mu'=8.1, \mu_a=4.05$,
in the right plot $\mu'=10, \mu_a=5$. The lowest two pink dotted lines is the homogeneous case with $\mu(x)=4.05$ and $\mu(x)=5$ respectively.
In all the plots we increase $\alpha$ from $0$ to $1$ with a step $\delta \alpha=0.05$. }\label{fig2}
\end{center}
\end{figure}

{\it Interplay of disorder effect and periodic effect.-}
Holographic superconductor with periodic chemical potential has been studied in \cite{p1,p2,p3}.
In \cite{p2,p3} the authors found that the superconductivity is enhanced by the presence of the periodic chemical potential.
In Fig. \ref{fig2} we plot the average value of the order parameter $\langle \mathcal O_a \rangle$ as a function of $\alpha$ for
various combinations of $k_1$ and $k_2$ with fixed $\mu'$ and $\mu_a$.
The lowest pink lines in Fig. \ref{fig2} are the homogeneous solutions with $\mu(x)=\mu_a=4.05$ and 5.
It is known that $\mu_c=4.06$ is the critical value
for the homogeneous configuration after $\mu_c$ we will see no superconductivity \cite{h1}.
From the left plot it can be seen that $\langle \mathcal O_a \rangle=0$ when $\mu(x)=4.05<4.06$, while for the periodic or quasi-periodic cases
we have non-zero condensation for some regions of $\alpha$.
Similar phenomena also happen for the case of $\mu_a=5$. In all cases,
both periodic and quasi-periodic chemical potential induce a larger value of
order parameter compared to the homogeneous case.

When $\alpha=0$ or $\alpha=1$, we recover the cases of periodic chemical potentials: $\mu(x)=\mu_a+(\mu'-\mu_a)\cos(k_1\pi x) $ and $\mu(x)=\mu_a+(\mu'-\mu_a)\cos(k_2\pi x) $. From Fig. \ref{fig2}, we can see that $\langle \mathcal O_a \rangle$ decreases
with increasing $k$ in the periodic cases. As a check we see that when $k=k_2=7/2; 9/2$, which is greater than $k_1=2$, $\langle \mathcal O_a \rangle$ for a periodic $\mu(x)$ with $k=k_2$ is small than that of $k=k_1$.
If we keep increasing $k$ (the results are not include here), the condensation $\langle \mathcal O_a \rangle$ asymptotes some constant value.
These results have also been found in \cite{p1,p3}.

Looking at the two red lines with dots in the top of Fig. \ref{fig2}, we see the condensation does not monotonically increase with increasing $\alpha$. The condensation decreases first then increases when we increase the portion of the
case of $k_2=3/2$ by tuning $\alpha$. This means that there is an interplay between the disorder effect and the periodic effect: periodic chemical potential favors an
increasing condensation, while disorder favors a decreasing one.
In the left plot of Fig. \ref{fig2}. Similar non-monotonic behaviors of the
two cases with $k_2=7/2$ (blue lines) also confirm the existence of disorder effect.
we can also see a phase transition from the superconducting phase to
a normal phase at $\alpha_c\sim 0.8$ when $k_1=2, k_2=9/2$, but the main reason
of the phase transition is the periodic effect since the transition happens at $\alpha_c>0.5$ and the
periodic case with $k=k_2=9/2$ is of a vanishing condensation.

\begin{figure}
\begin{center}
\includegraphics[trim=0cm 10cm 0cm 10cm, clip=true,scale=0.38]{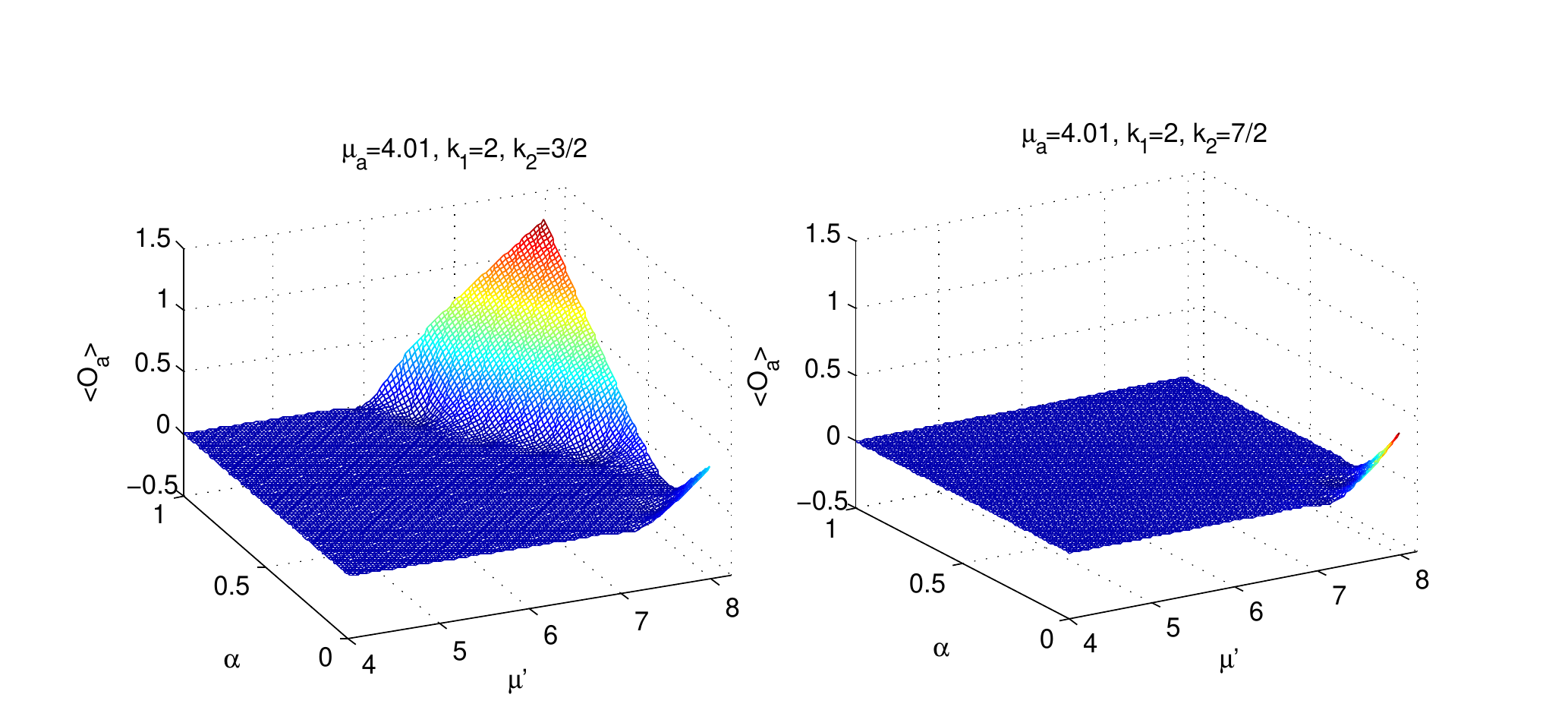}
\includegraphics[trim=0cm 10cm 0cm 10cm, clip=true,scale=0.45]{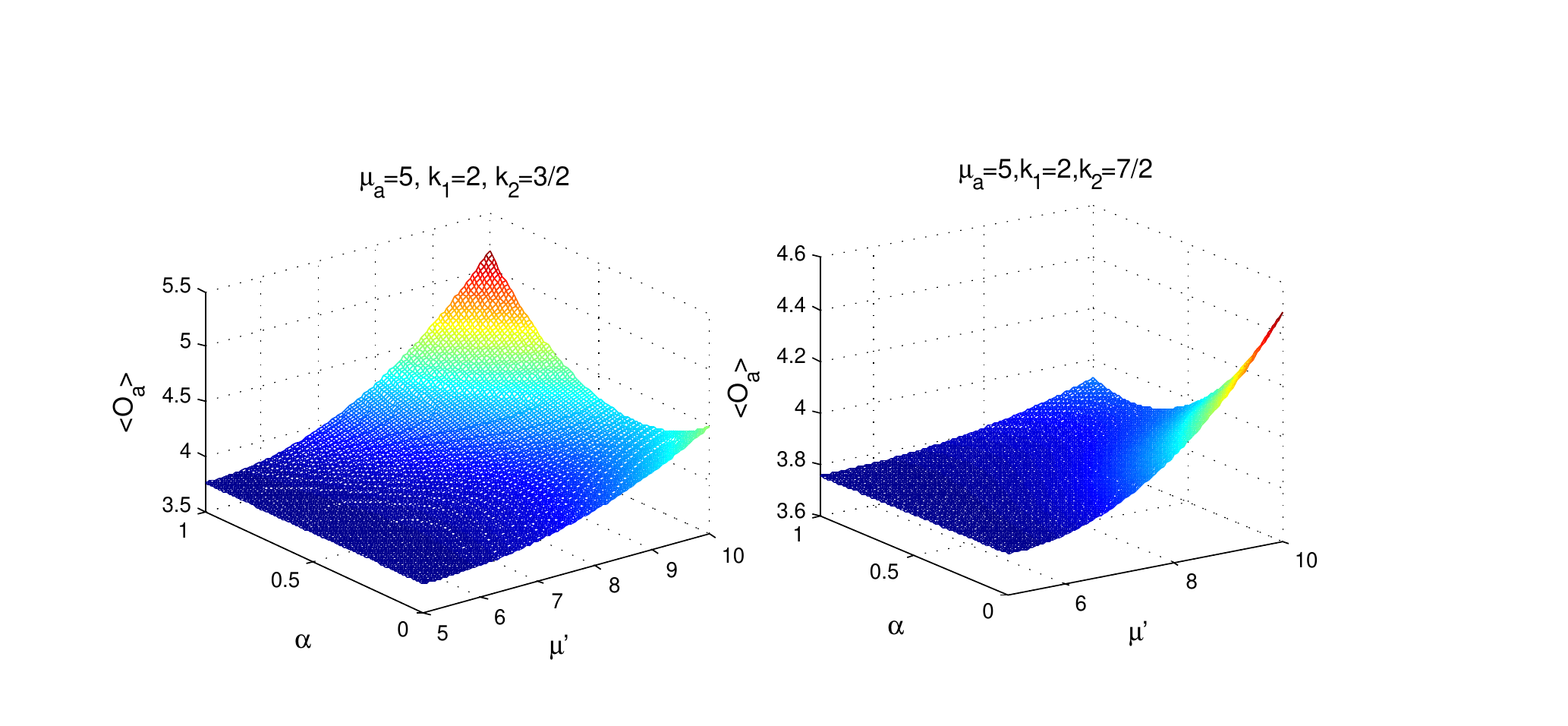}
\caption{$\langle \mathcal O_a \rangle$ as a function of both $\alpha$ and $\mu'$ for a fixed $\mu_a=4.01$ and $\mu_a=5$,
 there are regions in which the condensation is zero when $\mu_a=4.01$, which means that there is phase transition when increasing $\alpha$ when $7.1\preceq\mu'<8.02$.}\label{fig3}
\end{center}
\end{figure}

We also studied
how the condensation behaves when we tune both $\mu'$ and $\alpha$ with a fixed
$\mu_a$. Figure \ref{fig3} shows
$\langle \mathcal O_a \rangle$ as a function of both $\alpha$ and $\mu'$, where $0<\mu'< 2 \mu_a$ is chosen in order to have positive chemical potentials.
The important information from Fig. \ref{fig3} is that when we reduce $\mu'$ (the oscillating amplitude) with fixed $\alpha, \mu_a, k_1$ and $k_2$, the condensation will be decreased.

The two parameters $\alpha$ and $\mu'$
control the properties of the disorder effect, and
the quasi-periodic $\mu(x)$ affect the superconductor in a complex way.
With a fixed $\alpha$, increasing the amplitude $2(\mu'-\mu_a)$ of $\mu(x)$ enhances the superconductivity, as shown in Fig. \ref{fig3}.
When $\alpha=$ 0 or 1 we reproduce the result that the superconductivity of a striped holographic superconductor
will be enhanced \cite{p2,p3,p4}.

However, with a fixed amplitude $2(\mu'-\mu_a)$, the disorder can always
suppress the superconductivity when by turning $\alpha$ from zero to a finite value,
as shown in Fig. 1, Fig. 2, Fig. 3 and Fig. 4.

The interplay between the disorder effect and the periodic effect with fixed $\mu'$ and $\mu_a$ will result in a phase transition from the superconducting state to a non-superconducting state in some regions
of parameters as shown in Fig. 1 and Fig. 3. The DC conductivity along the $y$ direction of the non-superconducting state is finite, as shown by the inset in Fig. 1 ($\alpha=0.25$), which means that the non-superconducting state is a normal metal state rather than an insulating state.

{\it Discontinuous phase transition from superconducting to normal state.-}
With the results in the above section,  we already see that there is an phase transition
when the superconductor is close to $T_c$ ($\mu_a \approx \mu_c=4.06$) by increasing $\alpha$
form zero to a finite value (<0.5) for $\mu_a=4.02$ and $\mu_a=4.01$  as shown in Fig. 1 and Fig. 3 .
Figure 4 shows the critical value of $\alpha_c< 0.5$,
at which a phase transition from the
superconducting state to the normal state occurs when $\mu_a=4.02$ and $\mu_a=4.01$.
We note that the value of $\alpha_c$ for the case of $\mu'=8.02, \mu_a=4.01, k_1=2, k_2=3/2$ ($\simeq 0.2$) is larger than that for the case of $\mu'=8.02, \mu_a=4.01, k_1=2, k_2=7/2$, ($\simeq 0.15$), which is a consequence of the interplay between the disorder effect and the periodic effect as studied above.
From the four insets in Fig. 4 (blue lines), we see the order parameter goes discontinuous at $\alpha_c$, which indicates that the
superconducting to normal phase transition is a discontinuous one.
By computing many cases with other values of $\mu_a$ systematically, we find that when $\mu_a>4.03$ there is no disorder
driven phase transition anymore with $k_1=2$.

When the phase transition happens, the free energy of the
superconductor is also obtained
by computing the on-shell action according to the AdS/CFT dictionary.
The results are shown in Fig. 5.
It is clearly to see that the free energy also goes discontinuously at $\alpha_c$, which means that this is a zeroth order phase transition. More details for the calculations of conductivity and free energy will be presented elsewhere \cite{dzeng}.



\begin{figure}
\begin{center}
\includegraphics[trim=1cm 9.5cm 3cm 9.5cm, clip=true,scale=0.55]{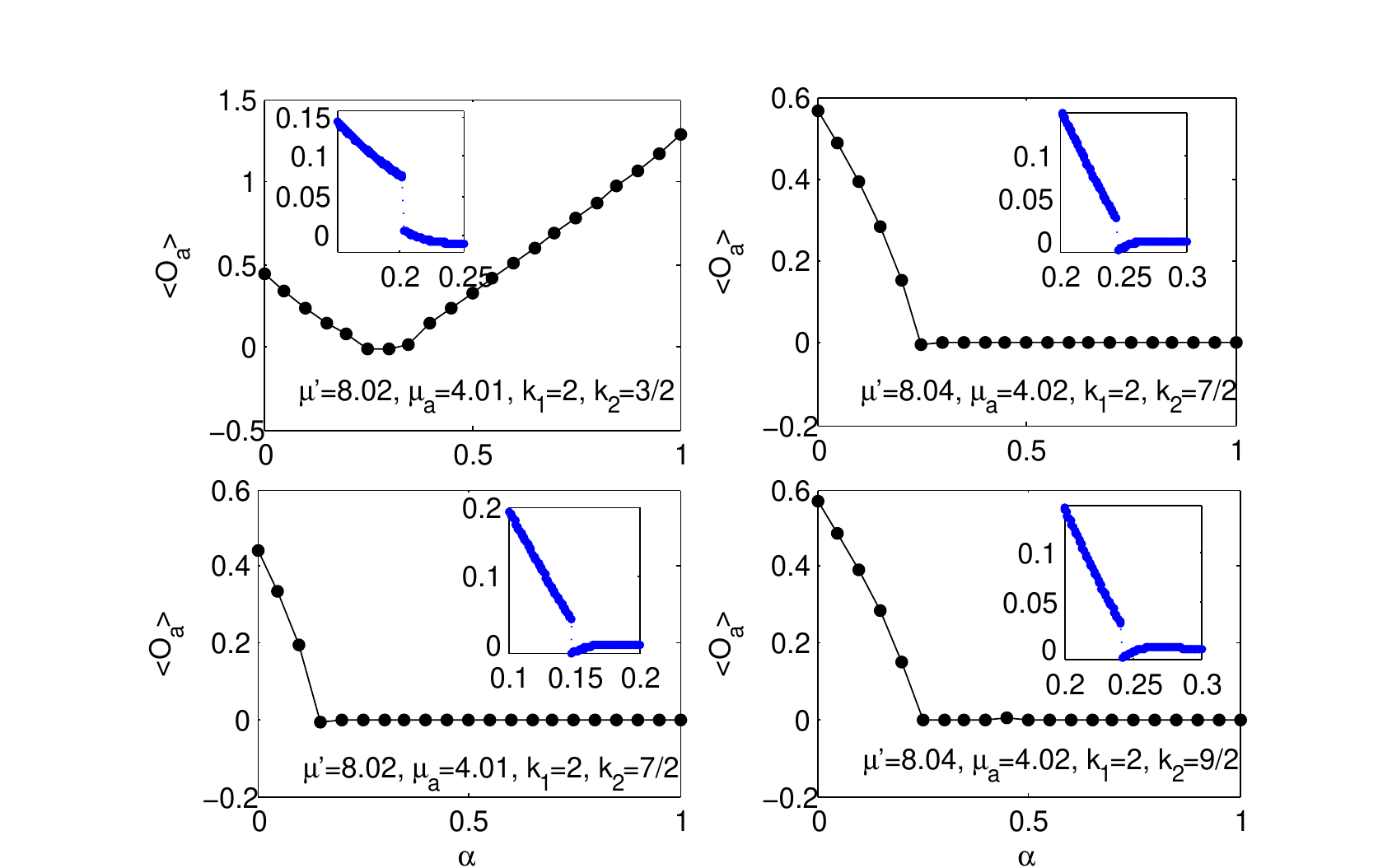}
\caption{The Anderson localization phase transition with $\mu_a=4.01; \mu'=8.02$ and $\mu_a=4.02; \mu'=8.04$.
The four insets are the detail plot of the region when phase transition happens, in  which
we increase $\alpha$ step by step with the distance $\delta \alpha=0.001$.   }\label{fig4}
\end{center}
\end{figure}

\begin{figure}
\begin{center}
\includegraphics[trim=0cm 10cm 6cm 10cm, clip=true,scale=0.55]{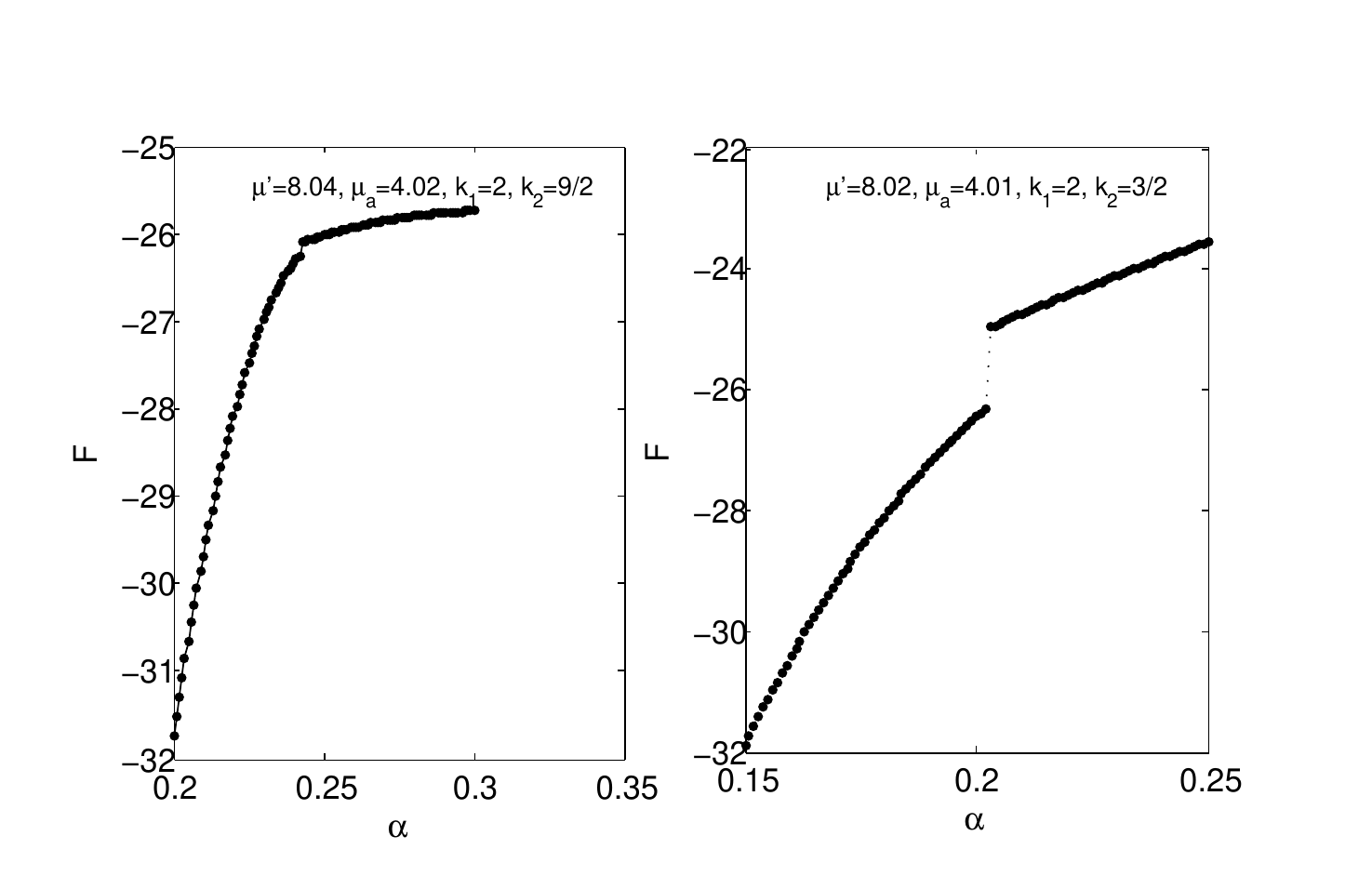}
\caption{The free energy of the superconductor when the phase transition happens.
(left) $\mu'=8.04, \mu_a=4.02, k_1=2, k_2=9/2$.(right)$\mu'=8.02, \mu_a=4.01, k_1=2, k_2=3/2$. }\label{fig5}
\end{center}
\end{figure}

{\it Conclusion.-}
In this paper, we systematically studied the interplay of disorder effect and periodic effect in two dimensional
$s$-wave holographic superconductors. We reproduced the results in condensed matter physics that the disorder will suppress superconductivity
and finally result in a discontinuous superconducting to normal state phase
transition when the gap is sufficiently small relative to the strength of disorder.
It seems natural to interpret the phenomena as an Anderson localization in holographic 
superconductor, however we are working in the probe limit and there is no momentum dissipation,
and we do not see the signal of a transition from the normal state to the insulating state when increasing the disorder for $T>T_c$.
Then it is still need further studying to make sure if we can say this is a holographic realization in AdS/CFT.

\textbf{Acknowledgments} We thank Zhe Yong Fan, Li Li, D. Arean and Sean Hartnoll for many valuable comments.
We especially thank Antonio M. Garc\'{\i}a-Garc\'{\i}a suggested me to study the quasi-periodic lattice effect, we especially thank
Hai Qing Zhang for discussing the numerical method and the periodic cases.
HBZ is supported by the National Natural Science Foundation of China (under
Grant No. 11205020). HBZ is also partly support from a FCT, grant PTDC/FIS/111348/2009 and a Marie Curie International Reintegration Grant
PIRG07-GA-2010-268172.


\vspace{-5mm}
\begin{thebibliography}{99}
\vspace{-5mm}

\bibitem{1}J. Bardeen, L. N. Cooper, and J. R. Schrieffer, Phys. Rev.
{\bf 108}, 1175¨C1204 (1957).
\bibitem{2} P. Anderson, Journal of Physics and Chemistry of Solids, {\bf vol.11}, 26.

\bibitem{3} M. Ma, P.A. Lee, Phys. Rev. B {\bf32}, 5658, (1985).
\bibitem{31}  A. M. Goldman and N. Markovi¡äc, Physics Today {\bf51} ,
(11), 39 (1998).
\bibitem{32} Y. Dubi, Y. Meir, Y. Avishai, Nature {\bf449}, 876 (2007).

\bibitem{33} E. Nakhmedov, R. Oppermann,Phys. Rev. B {\bf81}, 134511 (2010).


\bibitem{35} A. M. Finkelstein, Physica B {\bf197}, 636 (1994).

\bibitem{36} P. W. Anderson, K. A. Muttalib and T. V. Ramakrishnan, Phys. Rev. B {\bf28}, 117 (1983).

\bibitem{37} A. Kapitulnik and G. Kotliar, Phys. Rev. Lett. {\bf54}, 473 (1985).

\bibitem{38} M. Ma and Eduardo Fradkin, Phys. Rev. Lett. {\bf56}, 1416 (1986).

\bibitem{4} S. A. Hartnoll and C. P. Herzog, Phys. Rev. D {\bf77} 106009 (2008)
[arXiv:0801.1693 [hep-th]].
\bibitem{9} M. Fujita, Y. Hikida, S. Ryu and T. Takayanagi, JHEP {\bf0812}
(2008) 065 [arXiv:0810.5394 [hep-th]].
\bibitem{5} S. Ryu, T. Takayanagi and T. Ugajin, JHEP {\bf1104} (2011) 115
[arXiv:1103.6068 [hep-th]].
\bibitem{6} A. Adams and S. Yaida, arXiv:1102.2892 [hep-th].
\bibitem{7} A. Adams and S. Yaida, arXiv:1201.6366 [hep-th].
\bibitem{8} O. Saremi, arXiv:1206.1856 [hep-th].

\bibitem{adscft}
  J.~M.~Maldacena,
  Adv.\ Theor.\ Math.\ Phys.\  {\bf 2} (1998) 231;
  S.~S.~Gubser, et al.,
  Phys.\ Lett.\  B {\bf 428} (1998) 105;
  E.~Witten,
  Adv.\ Theor.\ Math.\ Phys.\  {\bf 2} (1998) 253.


\bibitem{h1}
  S.~S.~Gubser,
  Phys.\ Rev.\  D {\bf 78} (2008) 065034;S.~A.~Hartnoll, C.~P.~Herzog and G.~T.~Horowitz,
  Phys.\ Rev.\ Lett.\  {\bf 101} (2008) 031601.

\bibitem{h2}G. T. Horowitz, J. E. Santos, B. Way, Phys. Rev. Lett. {\bf106}, 221601 (2011);
M. Montull, A. Pomarol, Pedro J. Silva, Phys. Rev. Lett. {\bf103}, 091601 (2009);  M. J. Bhaseen et. al, Phys. Rev. Lett. {\bf110} 015301 (2013).

\bibitem{d1} D. Arean, A Farahi, L. A. P. Zayas, I. S. Landea, A. Scardicchio,	 arXiv:1308.1920 [hep-th]

\bibitem{qp1}X. Cai, L.J. Lang, S. Chen, Y. Wang, Phys. Rev. Lett. 110, 176403 (2013).
\bibitem{qp2}Roati, G. et al., Nature {\bf453}, 895-898 (2008).
\bibitem{qp3}M. Tezuka and A. M. Garcia-Garcia, Phys. Rev. A {\bf82}, 043613 (2010).
\bibitem{spectral}L. N. Trefethen,{\it Spectral methods in MATLAB}, Siam, Philadelphia, (2000).

\bibitem{p1} R. Flauger, E. Pajer and S. Papanikolaou, Phys. Rev. D {\bf83}
 064009 (2011)[arXiv:1010.1775 [hep-th]].

\bibitem{p2} J. Erdmenger, X. -H. Ge and D. -W. Pang, arXiv:1307.4609
[hep-th].

\bibitem{p3} S. Ganguli, J. A. Hutasoit, G. Siopsis,
 Phys. Rev. D {\bf86}, 125005 (2012).

\bibitem{p4} G. T. Horowitz and J. E. Santos,  [arXiv:1302.6586 [hep-th]].

\bibitem{dzeng} Hua Bi Zeng, in preparation.

\end{thebibliography}
\end{document}